\documentclass{ws-procs9x6-cpt22}
\begin{document}

\newcommand{\refeq}[1]{(\ref{#1})}
\def\etal {{\it et al.}}
\newcommand{\aegis}{AE$\bar{\hbox{g}}$IS}

\title{Antimatter Research: Advances of \aegis{} }

\author{B. \ Rien\"acker,$^1$}

\address{$^1$Physics Department, University of Liverpool,\\
Liverpool L69 3BX, UK}

\author{On behalf of the \aegis{} Collaboration}

\begin{abstract}
The \aegis{} collaboration is underway to directly measure the gravitational free-fall of neutral antimatter atoms. The experiment recently succeded in producing a pulsed cold antihydrogen source for the first time, and has now entered into its second phase, which aims at the formation of a slow antihydrogen beam and subseuqently a first proof-of-concept gravitational measurement. Major upgrades have been made, such as an improved antihydrogen production scheme and a new state of the art antiproton trap. \aegis{} was also connected to CERN's new antiproton deceleration facility ELENA and achieved first antiproton catching in late 2021. 
\end{abstract}

\bodymatter

\section{Introduction} \label{intro}
As evident as the missing antimatter half of the universe may seem, there still has been no experimental observation of a process that could explain this. The antimatter experiment \aegis{}, located at the Antimatter Factory at CERN, aims at using a pulsed beam of antihydrogen atoms to directly measure the effect of gravity on these neutral antimatter particles \cite{aegis01}. Any abnormal behavior would indicate a violation of the Weak Equivalence Principle (WEP) and hint at new physics.
\aegis{} exploits the charge exchange reaction between cold trapped antiprotons and Rydberg-excited positronium (Ps) in order to create a pulse of antihydrogen with high efficiency. The figure of merit using this reaction type rather than more direct methods such as radiative recombination is the scaling of the antihydrogen formation cross-section \cite{aegis02} with $\text{n}_{\text{Ps}}^4$ and the increase of the Ps lifetime in Rydberg-levels, which scales with the third power or the principal quantum number $\text{n}_\text{Ps}$. Using the knowledge of the antihydrogen formation time together with the micrometer-precise vertical displacement of the beam due to Earth's gravitational influence will yield a directly obtained value for the gravitational acceleration of antimatter.

\section{Current state and results}\label{section1}
In \aegis{} first phase, which lasted until 2019, a successful demonstration of a pulsed source of antihydrogen was achieved \cite{aegis03}. The main ingredients to this result were: a positron accumulation system based on a originally $50\,\text{mCi}$ intense positron source ($^{22}$Na) and two consecutive Surko-style buffer gas traps. The system provided every few minutes a pulse of several million positrons; a nanostructured porous silicon target with the ability to provide Ps in the range of environmental tempatures \cite{aegis04}; a highly specialized laser system enabling positronium excitation via two-step optical transitions, firstly from the $1^3S\rightarrow3^3P$ states and secondly from $3^3P\rightarrow$ Rydberg-levels with wavelength tunable nanosecond-precise laser pulses \cite{aegis05}; a with electrons sympathetically cooled and trapped antiproton plasma below $440\,$K; finally a complementary detector setup based on a microchannel plate and several scintillator-photomultiplier tube assemblies.

A first important result of \aegis{} was the adoption of a Moire-deflectometer (the classical counterpart to a Mach-Zehnder interferometer) for antimatter. Antiprotons from the AD were degraded to several keV and shot through a couple of gratings with $12\,\mu$m wide slits and $40\,\mu$m pitch in a distance of $25\,$mm from each other. Antiprotons were detected with a nuclear emulsion detector \cite{aegis06}. Comparing the obtained antiproton stripe pattern with a light reference, it was established that the device maintains its functionality in the cryogenic vacuum of the experiment, and that forces down to $10^{-18}\,$N can be measured with the given seetup and particle velocities. By increasing the dimension of the deflectometer and reducing the final energy of produced antihydrogen (i.e. reaching enviromental temperatures), the required force range to detect effects of gravity is reached.

Another important outcome was the controlled Rydberg-Ps production in the $1\,$T magnetic field of the experiment. The maximum reachable Rydberg-level for Ps was n$_{\text{Ps}} = 18$, since for higher levels significant losses due to field-ionization occurred \cite{aegis07}. This resulted from a self-induced electric field experienced by the excited Ps atoms moving perpendicular to the magnetic field with a velocity v$_{\text{Ps}}$ of the order of $10^5\,\text{m\,s}^{-1}$. This is the so-called motional Stark effect, causing a major limitation to the chosen experimental scheme.

Finally, a first pulsed cold antihydrogen source was realized by letting positronium atoms with n$_{\text{Ps}} \approx 17$ interact with a plasma of $\approx 10^6$ antiprotons, kept inside a Malmberg-Penning trap with an opening at the top covered with a mesh grid to let Rydberg-Ps enter \cite{aegis03}. Antihdyrogen production was confirmed by comparing antiproton annihilation rates in the few $\mu$s after positron implantation into the converter target, permutatively with and without the presence of antiprotons, positrons and the lasers. A significant excess signal was observed when using antiprotons, positrons and laser at the same time, corresponding to a production rate of 0.05 antihydrogen atoms per cycle of the experiment (lasting for about $110\,$s).

\section{Advances and goals in the near future}\label{section2}
After the successful development of the first pulsed cold antihdyrogen source, the \aegis{} collaboration
launched its second major phase, with the goal to create a beam of antihydrogen and to achieve a first proof-of-concept inertial measurement. The main challenges identified during the first phase are the following:
\begin{itemize} 
	\item[-] The antihydrogen production rate has to be increased by at least two orders of magnitude.
	\item[-] The temperature of the antiproton plasma and also of the resulting antihydrogen atoms has to be reduced by one order of magnitude.
	\item[-] The experiment has to take place in the most homogeneous magnetic field region of the experiment for a proper free-fall measurement.
	\item[-] Laser pulses for positronium excitation cannot not remain inside the experiment, but have to be guided out again by using mirrors in order to avoid heating/systematic background signals.
	\item[-] A method to avoid field-ionization by self-induced motional Stark electric fields must be developed.
\end{itemize}
	
These challenges required a major upgrade of the experiment, in particular on the antiproton and positronium side. In response to that, a collinear production scheme was adopted, where the positron/Ps converter target is installed on the same axis as that of a new antiproton trap.  As a consequence, the motional Stark effect is cancelled at first order in the trajectory angle
$\theta$ against the central axis, because the mean Ps emission trajectory is now aligned with that axis. This will put us in the position to increase the Ps Rydberg level as high as n$_{\text{Ps}} = 32$, being limited only by the second order angular spread of the spatial Ps distribution, which is required to efficiently cover the entire antiproton plasma in the trap. An increase of charge exchange cross-section by one order of magnitude is thus expected. 
Another important improvement comes from the connection of \aegis{} to the new ELENA storage ring, which yields an increase of two to three orders of magnitude in trapable antiprotons, following directly from the reduced antiproton energy of $100\,$keV with respect to the $5.3\,$MeV antiprotons from the Antiproton Decelerator \cite{aegis08}.
Finally, the antihydrogen production trap of the experiment was fully redesigned in order to reduce the number of electrodes and free up roughly $20\,$cm of flight path for the antihydrogen atoms. In addition, the trap now is really cylindrically symmetric and is equipped with an active alignment system working at cryogenic temperatures, which will considerably enhance the plasma stability and liftime. Thus, we expect  plasma temperatures of few tens of Kelvin only.
In combination, all these upgrades will boost the antihydrogen production rate, yielding 1--10 atoms per single experimental cycle, with the sole disadvantage of a slightly cut view further downstream due to the Ps target in the center of the electrode. 

With its upgraded hardware and software, \aegis{} will soon follow its main goal to demonstrate the formation of a neutral antimatter beam and to send this beam through a set of matter gratings for inertial measurements and thus directly testing the Weak Equivalence Principle for antimatter.

\section*{Acknowledgments}
This work was sponsored by the European’s Union Horizon 2020 research and innovation program under the Marie Sklodowska Curie grant agreement No. 754496, FELLINI, by the Wolfgang Gentner Programme of the German Federal Ministry of Education and Research (grant no. 05E18CHA and by Warsaw University of Technology within the Excellence Initiative: Research University (IDUB) programme and the IDUB-POB-FWEiTE-1 project grant.

\end{document}